\documentclass{emulateapj}
\usepackage{graphicx}
\usepackage{hyperref}

\begin{document}
\shorttitle{Inflow \& outflow with conduction for Sgr A*} \shortauthors{Shcherbakov R., Baganoff,
F.}

\title{Inflow-Outflow Model with Conduction and Self-Consistent Feeding for Sgr A*}   

\author{Roman V. Shcherbakov\altaffilmark{1}, Frederick K. Baganoff\altaffilmark{2}}
\altaffiltext{1}{\url{http://www.cfa.harvard.edu/\~rshcherb/}    Harvard-Smithsonian Center for Astrophysics, 60 Garden Street, Cambridge, MA
02138, USA} \altaffiltext{2}{Center for Space Research, Massachusetts Institute of Technology,
Cambridge, MA, 02139} \email{rshcherbakov@cfa.harvard.edu}

\begin{abstract}
We propose a two-temperature radial inflow-outflow model near Sgr A* with self-consistent feeding
and conduction. Stellar winds from individual stars are considered to find the rates of mass
injection and energy injection. These source terms help to partially eliminate the boundary
conditions on the inflow. Electron thermal conduction is crucial for inhibiting the accretion.
Energy diffuses out from several gravitational radii, unbinding more gas at several arcseconds and
limiting the accretion rate to $<1\%$ of Bondi rate. We successfully fit the X-Ray surface
brightness profile found from the extensive Chandra observations and reveal the X-Ray point source
in the center. The super-resolution technique allows us to infer the presence and estimate the
unabsorbed luminosity $L\approx4\cdot10^{32}{\rm erg~s^{-1}}$ of the point source. The employed
relativistic heat capacity and direct heating of electrons naturally lead to low electron
temperature $T_e\approx 4\cdot10^{10}$~K near the black hole. Within the same model we fit $86$~GHz
optically thick emission and obtain the order of magnitude agreement of Faraday rotation measure,
thus achieving a single accretion model suitable at all radii.
\end{abstract}


\section{INTRODUCTION}
Our Galaxy hosts a supermassive black hole (BH) with a mass $M=4.5\cdot10^6 M_\odot$
\citep{ghez,reid} at a distance $R=8.4$~kpc. The BH exhibits low luminosity state probably due to
inefficient feeding and cooling. Almost all available matter outflows from the region, whereas only
the small fraction accretes \citep{quataert_wind}. This feeding region within several arcseconds
contains X-Ray emitting gas, but some X-Rays are expected from a synchrotron self-Compton (SSC) or
synchrotron source from accretion at several Schwarzschild radii $r_{\rm g}$. The study of X-Rays
offers a unique opportunity to test the full range of accretion scales from several $''$ to $r_{\rm
g}=10^{-5}$$''$ and construct a single model.

Modeling the accretion flow with such a huge range of scales is a challenge.  3D SPH simulations
are performed in the outer region between $1''$ and $10''$ \citep{rockefeller,cuadra}. Latest MHD
simulations \citep{sharma_spherical} are limited to $3$ orders of magnitude in radius and axial
symmetry. Only the one-dimensional calculation \citep{quataert_wind} can in principle resolve the
flow everywhere. Thus, 1D modeling is the approach we adopt extending it down to the BH horizon.

We analyze the quiescent observations \citep{muno_diff} of X-Ray emission from central several
arcseconds around Sgr A* in \S \ref{s_obs}. The total exposure is 25 times longer compared to
previously analyzed data \citep{baganoff}. The super-resolution processing based on spacecraft
dithering helps resolving sub-pixel scales. The up-to-date data on stellar wind emitters are
summarized in \S \ref{s_winds}. We smooth matter ejection rates of individual stars over radius and
sum them into a single feeding rate, also properly averaging the wind velocity. This presents a
significant improvement over an ad-hoc feeding in \citet{quataert_wind}. The dynamical
two-temperature equations are derived in \S \ref{s_dynam}. We consider the electron conduction the
main energy transport mechanism, approximating the unsaturated heat flux by a simple formula. The
Bondi flow \citep{bondi52} without heat transport overestimates the X-Ray luminosity by a factor of
$10^3.$ The other important effects considered are the relativistic heat capacity of electrons and
superadiabatic heating equivalent to entropy production. The ways to solve the resulting system of
equations and corresponding results are presented in \S \ref{s_sol}. We employ the shooting method
and find the minimum $\chi^2$ fit for X-Ray surface brightness profile, simultaneously fitting
$86$~GHz flux. The best fit model requires X-Ray point source. The viability of a non-cooling
radial flow is examined.
\section{OBSERVATIONS}\label{s_obs}
\begin{figure}[h]
\includegraphics [scale=0.8,bb= 20 -10 275 375]{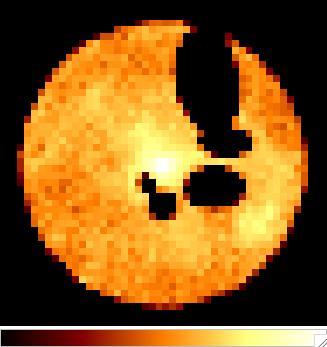}
 \caption{Chandra image of central 6'' around Sgr A*. Point sources and strong extended features are subtracted.}
 \label{fig_image}
\end{figure}

Central several arcseconds of the Galaxy were observed quite often over the past several years. The
rich region contains point sources identified as X-Ray binaries \citep{muno_point} and extended
emission features \citep{muno_diff} together with the source coincident with Sgr A*. The latter is
expected from hot accreting gas, and source confusion is practically impossible \citep{baganoff}.
Sgr A* source exhibits significant X-Ray flares associated with the SSC mechanism
\citep{baganoff_flares} or synchrotron \citep{katie}.  We are interested in quiescent emission, so
we exclude the flaring state. We bin the observations in $628$ seconds as a compromise between the
time resolution and the number of counts. About $4$ photons on average are received during $628$
seconds and we take only the observations with less than $15$ photons, thereby accumulating
$953$~ks in the quiescent state. The quiescent state also produces some point source X-Rays, likely
associated with SSC \citep{moscibr_sim}. We model these by a PSF-broadened central point source. We
eliminate the emission from the point sources and bright extended sources offset from Sgr A* (see
Figure~\ref{fig_image}). The bright extended emission may arise from the colliding winds of two
strong close emitters or from the collision of hot outflowing material with cold molecular
material. We exclude both effects from modeling of an averaged flow pattern.

\begin{figure}[h]
\includegraphics [scale=0.35, bb= 20 -10 175 575]{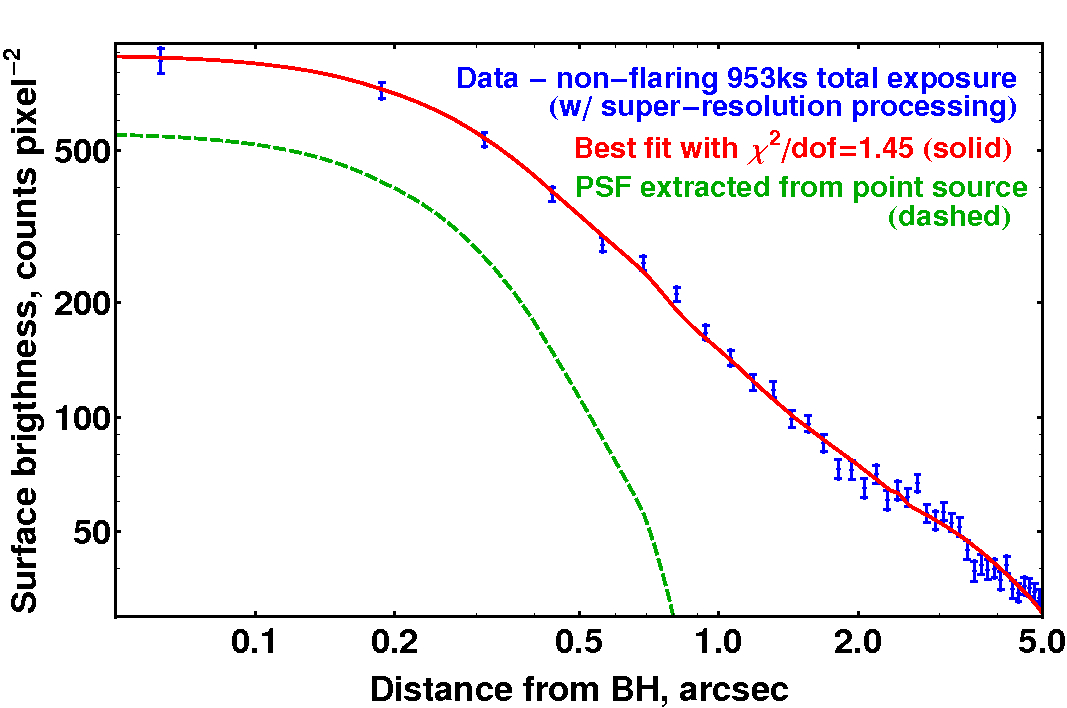}
\caption{Observed radial surface brightness profile (error bars), best
fit (solid) and the point source contribution to emission (dashed). The point source contribution is the scaled PSF.} \label{fig_model}
\end{figure}

We construct the surface brightness profile in counts per pixel squared for the duration of
observation as a function of distance from the BH. The size of Chandra pixel is $0.5'',$ which may
seem to pose a limit on radial binning of brightness profile. However, the position of satellite is
not steady over the duration of observations, but is findable with the $0.1''$ accuracy by
comparing with the known positions of bright point sources. Then we can achieve $0.1''$
super-resolution accuracy in surface brightness profile from knowing the orientation of the
detector pixels at any given time. The final profile is shown on Figure~\ref{fig_model} (error
bars) together with the point-spread function (PSF) (dashed) found from the nearby point source
J174540.9-290014 \citep{muno_point}. The PSF is scaled to match the contribution from the point
source. The counts cease to be monotonic at about $5''$ due probably to the production of X-Rays in
collisions of cold and hot regions. Therefore, only radiation within the central $5''$ is to be
modeled. As we are interested in how symmetric the surface brightness profile is, we divide the
emitting region into $4$ sectors $90\deg$ each centered on Sgr A* and extract the surface
brightness profile in each sector. The standard deviation of counts between sectors is below
$2\sigma$ the noise within $5'',$ but rises to several $\sigma$ outward from $5''.$ This justifies
our choice of the outer radiation boundary and proves the applicability of the radial model. Let us
now look in more details on manufacturing of the X-Ray emitting gas.
\section{STELLAR WINDS FEEDING}\label{s_winds}
The Galactic Center region has a concentration of massive Wolf-Rayet and blue giant stars,
expelling strong winds from their surfaces \citep{martins}. As the strongest wind emitters are
usually the brightest stars, all wind emitters are easily identifiable. We take the latest data on
ejection rates and velocities \citep{martins,cuadra} and complement them with the orbital
parameters of stars \citep{paumard,lu_ghez}. Following \citet{cuadra}, we minimize eccentricities
for the stars not belonging to the stellar disks as identified by \citet{lu_ghez}. The wind speeds
$v_w$ and ejection rates are taken directly from \citet{cuadra}.

There are several ways to treat the winds. \citet{rockefeller} performed a simulation with winds
from steady stars, whereas \citet{cuadra} considered moving stars. In both cases the time to reach
the quasi-steady solution $300-1000$~yrs is comparable to or longer than the orbital period at the
stagnation point $350$~yrs. Thus, it is reasonable to average over stellar orbits in a search for a
steady-state prescription of feeding. We reconstruct the full 3D orbits, but retain only the
apocenter and pericenter distances for the stars. We smooth the total wind ejection rate for each
star over the radial extent of its orbit and then smooth with the narrow Gaussian filter to
eliminate the divergences at the turning points.
 \begin{figure}[h]
\includegraphics [scale=0.51, bb= 20 -10 175 595]{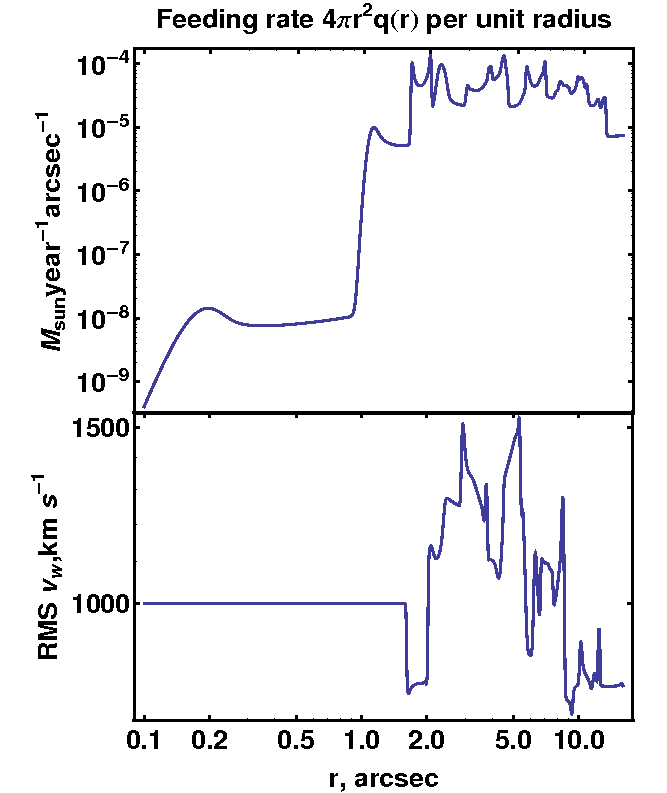}
 \caption{Mass input into the feeding region around the BH on the upper panel. Square averaged wind velocity $v_w$ on the lower panel.
Feeding is averaged over stellar orbits. Each wiggle represents a turning point of a single orbit.
Only S02 star feeds matter within $0.8$''.}
 \label{fig_feed}
\end{figure}
We add the resultant feeding profiles together to obtain the total feeding rate as a function of
radius (see Figure~\ref{fig_feed}). We square average the wind velocities weighing the contribution
of each star by its mass loss rate. However, the winds also acquire the velocity of a star as
viewed by a distant observer. We neglect stars' proper motions in calculations of wind energy. They
are negligible at several arcseconds, but would rather contribute to the angular velocity of matter
within $1'',$ where feeding is dominated by few stars. The dependence of the averaged wind speed on
radius is shown on Figure~\ref{fig_feed}. \citet{quataert_wind} assumed the power-law mass
injection rate $q(r)\propto r^{-\eta}$ for $r\in[2'',10''].$ The power-law index $\eta=2$
corresponds to zero slope of $\dot{M}(r)\propto r^2q(r)$ (see Figure~\ref{fig_feed}) and agrees
better with the present calculations, whereas their choice of constant wind velocity does not agree
with the present estimate.

We also incorporate S02 star \citep{martins_s2} into the calculations. The mass loss rate
$\dot{M}_{S02}=6\cdot10^{-8}M_\odot {\rm year}^{-1}$ of S02 is taken to coincide with that of
$\tau$ Sco. S02 has a spectral type $B0-2.5 V$ and a mass $M\approx16M_\odot$
\citep{mokiem,martins_s2}, whereas $\tau$ Sco has a very close type $B0.2V$ and a mass
$M\approx15M_\odot$ \citep{mokiem}. The inferred accretion rate onto the black hole
\citep{sharma_heating,sharma_rotation} $3\cdot10^{-8}M_\odot {\rm year}^{-1}$ is actually smaller
than $\dot{M}_{S02},$ thus the whole accreted material can in principle be provided by a single
weak wind emitter. This result is very different from \citet{cuadra}, who assumed all the matter
accretes from the inner boundary of the simulation, thus obtaining in a simplified treatment a much
larger accretion rate. However, the direct feeding mechanism \citep{loeb} by S02 does not work, as
its revised $\dot{M}_{S02}$ is much below the value required for feeding without the angular
momentum. In turn, the direct feeding by IRS 13E3 \citep{moscibr} produces too large accretion rate
in the absence of conduction.
\section{DYNAMICAL EQUATIONS}\label{s_dynam}
\subsection{Energy transport mechanism}
Radiatively inefficient flows can be mediated significantly by the energy transfer from the inner
regions to the outer \citep{blandford,johnson,sharma_spherical}. Such transfer happens in two
distinct ways: via convection or via diffusive energy transport. Convection is seen in numerical
simulations. It happens via Alfven instability \citep{igumen06} and magneto-thermal instability
(MTI) \citep{sharma_spherical} and modifies the density profile. Let us show that the electron heat
conduction wins over convection in the accretion flow. First, the MTI is driven by thermal
conduction, at any moment the electron conduction flux is larger then the MTI-induced heat flux.
Convection implies the motion of large-scale magnetized eddies, which in turn split into smaller
eddies and develop the whole turbulent cascade. In such settings the electron conduction is only
inhibited a factor of $\sim 5$ \citep{medvedev}. The speed of electrons is a factor of
$\sqrt{m_p/m_e}$ larger than the sound speed and the convection is subsonic; the same factor lowers
the ion diffusive heat transport. The relative strength of convective heat flux is proportional to
the gradient of logarithmic entropy, which is normally weaker than the proportionality to the
gradient of logarithmic temperature of conductive flux. Combining both effects we conclude that, if
there is convection or diffusion, then there is stronger conduction.  Severe inhibition of electron
conduction happens, if the turbulent cascade does not develop and mixing is absent. This is not the
case when the gas accretes. The strength of turbulent magnetic field increases then in the
convergent flow leading to dissipation and effective mixing \citep{shva,shcher}. It is reasonable
to think that the whole turbulent cascade develops and the electrons relatively freely find their
way around magnetic field lines to connect the different regions of the flow. When the electrons
and ions get decoupled from each other, the ion entropy may get equilibrated by convection, whereas
the electron temperature levels due to conduction. The investigation of this possibility is left
for future research. In present paper we take the energy transport to happen solely via electron
conduction.

There are several different regimes of conduction. First, the collisionality of the flow changes
from the large radii to the inner radii as the mean free path of particles $l$ exceeds the flow
size $r$. As the flow gets only weakly collisional at several arcseconds, the conductivity is well
approximated by a collisionless formula with $\kappa\propto r.$ Another assumption of the kind
deals with the electron velocity. As electrons can get only mildly relativistic, we take
conductivity to be proportional to square root of electron temperature $\kappa\propto \sqrt{T_e},$
instead of proportionality to relativistic electron velocity $\kappa\propto v_c$ \citep{johnson}.
When the gradient of electron temperature gets too large, the electrons transport heat via a
constant saturated flux, instead of the flux proportional to the gradient of temperature
\citep{cowie}. We check a posteriori that the flow is in an unsaturated heat flux regime. Finally,
we have for the heat flux $Q=-\kappa k_B dT_e/dr$
\begin{equation}\label{conductivity}
\kappa=0.1 \sqrt{k_B T_e/m_e}r n,
\end{equation} where $n=n_e$ is the electron density \citep{cowie}.

\subsection{System of Equations}
Gravitational energy of gas in the potential of an accretor is the ultimate inflow driver. It gets
transformed directly in several types: kinetic energy of bulk toroidal and radial motion, energy of
turbulent magnetic and velocity fields, thermal energy. Turbulent energy can also originate from
the toroidal shearing flow in a disk. Turbulence dissipates into thermal motions of ions and
electrons on the dynamical timescale, whereas ions and electrons exchange energy by slow Coulomb
collisions. The faster collective modes of ion-electron energy exchange may exist, though they may
not lead to equilibration of temperatures \citep{shkarofsky}. We do not separate the turbulent term
or write an equation on it for the purpose of current work, as its direct dynamical influence is
smaller than the influence of additional thermal energy produced via dissipation of turbulence and
entropy production \citep{shcher}. Following \citet{johnson}, we introduce the fractions $f_p$ and
$f_e$ of changes of gravitational energy, which go directly into thermal energy of ions and
electrons, but relate them via a direct heating mechanism \citep{sharma_heating}. For the purpose
of numerical stability we enhance Coulomb collisions by a factor of $1000,$ which effectively makes
ion and electron temperatures equal at large distances from the BH, but does not influence $T_e$
near the BH. Let us convert the qualitative ideas into equations.

 The composition of plasma determines the exact balance of the black hole gravitational pull
 and supporting gas pressure. Let us define the source function $q,$ so that the ejected mass of stellar winds
 per second is $\dot{M}_w=\int 4\pi~r^2~q~dr.$ We
denote the electron density by $n=n_e$ and write the continuity equation as
\begin{equation}
\frac{\partial n}{\partial t}+\frac1{r^2}\frac{\partial(n v_r r^2)}{\partial
r}=\frac{q(r)}{\mu_{av}},
\end{equation} where
\begin{equation}\mu_{av}\approx 1.14
\end{equation} is the average atomic mass per one electron for assumed solar abundance of fully ionized elements \citep{najarro}. The
ratio of number densities of atomic nuclei to electrons is
\begin{equation}
d=n_{\rm non-el}/n\approx 0.93.
\end{equation}

We write separate energy equations for electrons (e) and all ions (p) in terms of
\begin{equation}
c_{se}=\sqrt{\frac{k_B T_e}{m_p}} \quad{\rm and}\quad c_{sp}=\sqrt{\frac{k_B T_p}{m_p}},
\end{equation} assuming all ions have the same temperature. We set the speed of light equal unity $c=1$ and normalize to it all velocities.
 The ideal gas law gives normalized gas pressure
\begin{equation}
p_{\rm gas}=p_p+p_e=n (c_{se}^2+ d\cdot c_{sp}^2)
\end{equation} to be substituted into the Euler equation
\begin{equation}
\frac{D v_r}{D t}+\frac1{n \mu_{av}}\frac{\partial p_{\rm gas}}{\partial r}+\frac{r_{\rm
g}}{2(r-r_{\rm g})^2}+\frac{q(r)}{n \mu_{av}} v_r=0,
\end{equation} where $D/Dt=\partial/\partial t+v_r \partial/\partial r.$ The last term corresponds
to zero bulk radial velocity of emitted stellar winds.

The electron internal energy density can be approximated as
\begin{eqnarray}
u_e=m_e \left(\frac{3K_3(\theta_e^{-1})+K_1(\theta_e^{-1})}{4K_2(\theta_e^{-1})}-1\right)\approx
\\\nonumber\approx \frac32 \frac{0.7+2c_{se}^2m_p/m_e}{0.7+c_{se}^2m_p/m_e}m_p c_{se}^2.
\end{eqnarray} This takes into account the differential heat capacity of particles
\citep{shkarofsky}. The ion internal energy per particle is $u_p=3/2m_p c_{sp}^2.$

The energy exchange rate by Coulomb collisions is \citep{shkarofsky}
\begin{equation}\label{coulomb_energy}
F_{pe}=4.3\cdot10^{-19} \frac{n^2}{c_{se}^3}(c_{sp}^2-c_{se}^2).
\end{equation} The non-relativistic formula is used everywhere, as $F_{pe}$ rate is only significant in the
region of non-relativistic electrons. The energy equation for electrons is then
\begin{eqnarray}\label{energy_electr}
n\frac{D}{D t}\left(\frac32\frac{0.7+2c_{se}^2 m_p/m_e}{0.7+c_{se}^2 m_p/m_e}c_{se}^2\right)-c_{se}^2 \frac{D n}{D t}=C F_{pe}+\nonumber\\
-f_e n\frac{r_{\rm g}v_r}{2r^2}+\frac{q (1+d)}{2\mu_{av}}\left(\frac{v_r^2}{2}+\frac{v_w^2}{2}- \frac{5}{2}c_{se}^2\right)\\
+\frac1{r^2}\partial_r(r^2\kappa\partial_r c_{se}^2),\nonumber
\end{eqnarray} where $C\sim1000$ is the enhancement of collisions and conductivity is given by equation (\ref{conductivity}). The left-hand side
of the equation (\ref{energy_electr}) represents the compressive heating in the adiabatic flow. The
Paczynski-Wiita gravitational potential \citep{paczynski} is implemented for gravitational force,
but not in the entropy production term. This reflects the fact that the dissipation of turbulence
ceases near the BH as having slower timescale compared to the inflow time. The energy equation for
ions reads
\begin{eqnarray}\label{energy_prot}
n \frac{D}{Dt}\left(\frac32 c_{sp}^2\right)-c_{sp}^2\frac{Dn}{Dt}=-C F_{pe}\nonumber+\\
-f_p n\frac{r_{\rm g} v_r}{2r^2}+\frac{q(1+d)}{2\mu_{av}}\left(\frac{v_r^2}{2}+\frac{v_w^2}{2}-
\frac{5}{2}c_{sp}^2\right).
\end{eqnarray} The energy injection rate into ions is chosen to be the same per electron as
the energy injection rate into electrons to facilitate the equality of ion and electron
temperatures. Let us write a condition on $f_p$ and $f_e$ to decrease the number of free
parameters. We assume the ratio of heating fractions to be given by the direct heating mechanism
\citep{sharma_heating} as
\begin{equation}\label{heating_fraction}
\frac{f_e}{f_p}=\frac13\sqrt{\frac{T_e}{T_p}},
\end{equation} despite this calculation is non-relativistic and a large fraction of energy
dissipates at the small scales instead of direct large-scale heating.

\section{SOLUTIONS AND DISCUSSIONS}\label{s_sol}
We solve the derived system of equations from the outer boundary of the feeding region at
$14''=1.3\cdot10^6r_{\rm g}$ to the inner boundary at about $1.3 r_{\rm g},$ thus covering $6$
orders of magnitude in radius. Such a huge dynamic range requires the special solution technique,
the solution of a time-dependent system of equations \citep{quataert_wind} not being an option. We
employ the shooting method and find the smooth transonic solution through the inner sonic point at
$\sim3r_{\rm g}.$ In the presence of conduction the point, where sound speed equals inflow velocity
is not special anymore, and instead the point, where isothermal speed equals the inflow velocity,
plays the role of transonic surface \citep{johnson}. The system of equations is reduced to one
temperature in the outflow by setting $T_e=T_p$ and adding the equations (\ref{energy_electr}) and
(\ref{energy_prot}). The inner boundary is set at a point $r_{\rm in}$, where $dT_e/dr=0$ in a
non-conductive solution. Then for any non-zero conductivity the zero heat flux condition
$dT_e/dr=0$ is enforced at $r_{\rm in}.$ The outer boundary condition at $r_{\rm out}$ is
uncertain. It is natural to think the outflow would be transonic \citep{lamers}, however,
significant outer pressure may hold the gas in the subsonic regime near $r_{\rm out}.$ The position
of zero velocity stagnation point $r_{\rm st}$ determines the accretion rate $\dot{M}.$ Instead of
setting the pressure at the outer boundary we regulate that pressure by setting temperature $T_{\rm
st}$ at the stagnation point. Thus, we have 4 independent variables in the fit: accretion rate
$\dot{M},$ temperature at stagnation point $T_{\rm st},$ the ion heating rate $f_p$ and the
normalization $N$ of the point source contribution. They are all found iteratively to minimize
$\chi^2.$ We also iteratively find the positions of sonic point and stagnation point. The positions
of inner boundary and outer boundary are unchanged while solving the 4-point boundary value
problem.

The observed surface brightness radial profile is the data we fit. We generate a surface brightness
profile corresponding to the dynamical model by performing the optically thin ray tracing of X-Rays
at a set of photon energies and projected distances from the BH. We employ the up-to-date
bremsstrahlung emissivities (\citet{gould} and errata) and account for the emission by heavy
elements, excluding iron.  Solar metallicity interstellar absorption \citep{morrison} is assumed
with hydrogen column $N_{\rm H}=10^{23}$cm${}^{-2}.$ The fluxes are convolved with the response of
Chandra to find counts, then blurred with the energy-independent PSF (see Figure~\ref{fig_model})
and integrated over the radial extent of each bin.

The model with $\dot{M}=6\cdot10^{-8}M_\odot{\rm year}^{-1},$ $f_p=0.46,$ $T_{\rm
st}=3.2\cdot10^7$~K and $550{\rm counts~pixel}^{-2}$ produced at $r=0$ by a point source gives an
excellent fit with the minimum reduced $\chi^2=1.45$ and weighed $\chi^2_{\rm wei}=0.68$ with $1/r$
weights. The stagnation point is at $r_{\rm st}=1.01''.$ The correspondent unabsorbed point source
luminosity $L=4\cdot10^{32}{\rm erg~s}^{-1}$ is estimated for monoenergetic photons at $4$~keV and
agrees with the estimates of SSC luminosity in \citet{moscibr_sim}. Energy $4$~keV is chosen as the
energy Chandra is most sensitive to for assumed $N_{\rm H}.$ The minimum reduced $\chi^2=15$ is
achieved for the model without the point source. The models with the outer sonic point instead of
finite bounding pressure underpredict the X-ray surface brightness at several arcseconds, assuming
fixed $N_{\rm H}=10^{23}$cm${}^{-2}.$ The reliable fitting for $N_{\rm H}$ is possible only with
the use of spectral data and is left for future research. The assumption $T_p=T_e$ represents the
additional point of concern. Temperature equilibrium might not hold at the stagnation point at
$1''$ \citep{quataert_wind}, however the thermalization rate exceeds the outflow rate at $5''$ in
our subsonic dense outflow, thus $T_p=T_e$ holds there. The reliable modeling of non-equilibrium
flows requires the modeling of the whole spatial structure of the stellar winds and is left for the
future research as well.

\begin{figure}[h]
\includegraphics [scale=0.7, bb= 20 -10 175 495]{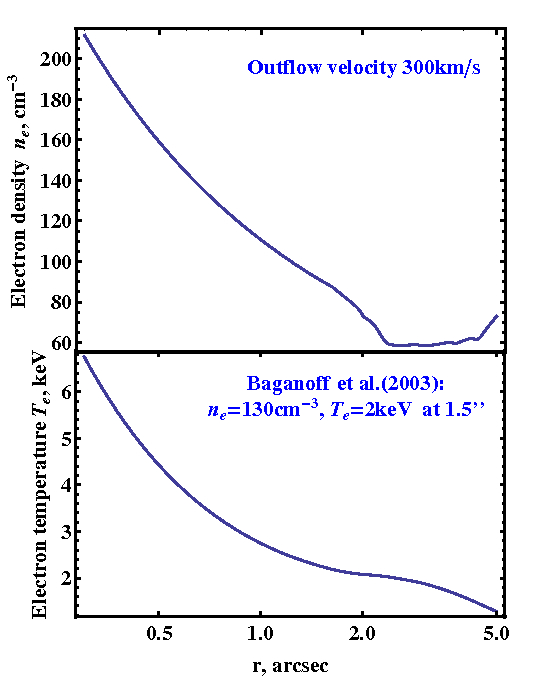}
\caption{Radial profiles of electron density $n=n_e$ in
cm~$^{-3}$ (upper panel) and electron temperature $T_e$ in keV (lower panel) in the feeding
region.} \label{fig_arcprof}
\end{figure}
The profiles of electron density $n_e$ and temperature $T_e$ within several arcseconds from the BH
are shown on Figure~\ref{fig_arcprof} and compare well with the simple earlier estimates
\citep{baganoff,quataert_wind}. The difference is that our best fit is a subsonic flow supported by
the outer medium with the density bounce at $5''.$ Though the achieved outflow velocity $v_{\rm
out}=300{\rm km~s}^{-1}$ is almost independent of radius for $r>2''.$ The line cooling
\citep{sutherland} reduces the heat contents only by several percent for gas reaching $5'',$
bremsstrahlung cooling being less important.

\begin{figure}[h]
\includegraphics [scale=0.7, bb= 20 -10 175 495]{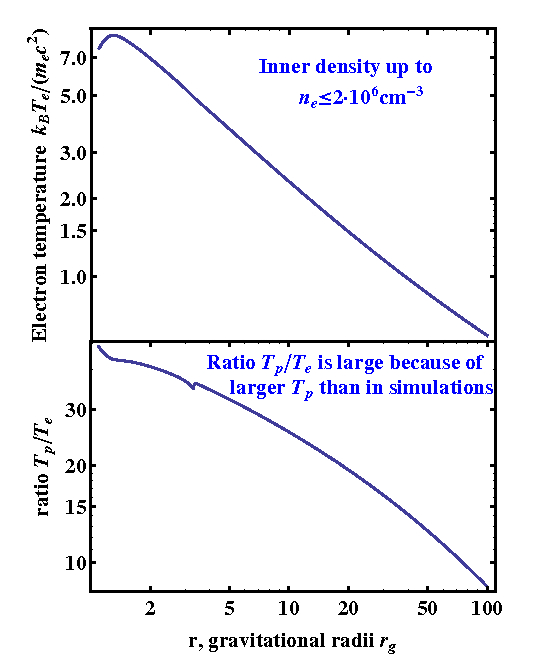}
\caption{Radial profiles of dimensionless electron temperature
normalized to electron mass $k_B T_e/(m_e c^2)$ (upper panel) and ratio of ion to electron
temperatures $T_p/T_e$ (lower panel) close to the BH. The inner sonic point is at $3r_{\rm g}.$} \label{fig_rgprof}
\end{figure}
The profiles of dimensionless electron temperature $k_B T_e/(m_e c^2)$ and ratio $T_p/T_e$ within
several Schwarzschild radii from the BH are shown on Figure~\ref{fig_rgprof}. The electron
temperature $T_e=4\cdot10^{10}$~K and density $n_e=2\cdot10^6{\rm cm}^{-3}$ are found close to the
BH. This dynamical model gives an excellent fit to the optically thick luminosity $L=1.73$~Jy at
$86$~GHz \citep{krichbaum} for assumed equipartition of thermal energy with the magnetic field. The
model overpredicts by a factor of $20$ the observed Faraday rotation measure $RM\sim 50{\rm
cm}^{-2}$ at $230$~GHz \citep{marrone}, but this may well be a geometric factor. The accretion
rate, temperature and density near the BH are in good agreement with more complicated models
specifically focusing on sub-mm emission \citep{sharma_spherical,moscibr_sim}. We notice that the
ratio of ion and electron temperatures $T_p/T_e$ is significantly larger than predicted by
\citet{moscibr_sim}, but probably because of the significantly lower $T_p$ in their numerical
simulations of the limited domain.

\section{Acknowledgements}
The authors are grateful to Ramesh Narayan for fruitful discussions, referee Eliot Quataert, Fu-Guo
Xie for encouraging us with the shooting method, Feng Yuan, Jorge Cuadra, Avi Loeb for
useful comments. The work is supported by NASA grants NNX08AX04H, NNX08AH32G, Chandra Award
GO9-0101X, SAO Award 2834-MIT-SAO-4018 and NSF grant AST-0805832.


\end{document}